\documentclass[epj]{webofc}
\usepackage[varg]{txfonts}   
\wocname{EPJ Web of Conferences}
\woctitle{ICNFP 2017}
\newcommand{\be}{\begin{equation}}
\newcommand{\ee}{\end{equation}}
\newcommand{\bea}{\begin{eqnarray}}
\newcommand{\eea}{\end{eqnarray}}
\begin{document}
\selectlanguage{english}
\title{Isospin dynamics in nuclear structure}
%
%

\author{Elena Litvinova\inst{1,2}\fnsep\thanks{\email{elena.litvinova@wmich.edu}} \and
        Caroline Robin\inst{3,4} \and
        Peter Schuck\inst{5,6}
}

\institute{Department of Physics, Western Michigan University, Kalamazoo, MI 49008, USA 
\and
           National Superconducting Cyclotron Laboratory, Michigan State University, East Lansing, MI 48824, USA 
\and
           Institute for Nuclear Theory, University of Washington, Seattle, WA 98195, USA 
\and
           JINA-CEE, Michigan State University, East Lansing, MI 48824, USA
\and
          Institut de Physique Nucl\'eaire, IN2P3-CNRS, Universit\'e Paris-Sud, F-91406 Orsay Cedex, France
\and
         Laboratoire de Physique et Mod\'elisation des Milieux Condens\'es, CNRS and Universit\'e Joseph Fourier, 25 Avenue des Martyrs, BP166, F-38042 Grenoble Cedex 9, France          
}

\abstract{%
 We discuss some special aspects of the nuclear many-body problem related to isospin transfer. The major quantity of interest is the in-medium propagator of a particle-hole configuration of the proton-neutron character, which determines the nuclear response to isospin transferring external fields. One of the most studied excitation modes is the Gamow-Teller resonance (GTR), which can, therefore, be used as a sensitive test for the theoretical approaches. Its low-energy part, which is responsible for the beta decay half-lives, is especially convenient for this. Models benchmarked against the GTR can be used to predict other, more exotic, excitations studied at nuclear rare isotope beam facilities and in astrophysics.
 As far as the precision is concerned, the major problem in such an analysis is to disentangle the effects related to the underlying interaction and those caused by the many-body correlations. Therefore, approaches (i) based on fundamental concepts for the nucleon-nucleon interaction which (ii) include complex many-body dynamics are the
preferred ones. We discuss progress and obstacles on the way to such approaches. 
}
\maketitle
\section{Introduction}
\label{intro}
The nuclear many-body problem, which undoubtedly remains among the most difficult problems in science, has been boosted tremendously during the last couple of decades. This can be related to the progress of the experimental rare isotope beam facilities, computational advancements as well as to the conceptual developments on the theory side. For few-body systems, the models based on various approaches to the bare nucleon-nucleon interaction, such as chiral effective field theory (EFT) \cite{ME.11}, modern Argonne \cite{Argonne}, Bonn \cite{CDBonn} and Nijmegen \cite{Nijmegen} potentials, are typically highlighted. Indeed, the combinations of such potentials with the advanced few-body techniques, in particular, no-core shell-model, coupled clusters theory, shell-model Monte-Carlo, in-medium similarity renormalization group and others, turn out to be rather successful in the description of the ground and low-energy excited states in light nuclei. However, these methods start to reveal deficiencies if they are applied to medium-heavy nuclear systems, possibly because of the appearance of the collective degrees of freedom. For the chiral EFT, in particular, introducing three-body forces 
has helped in moving up to the oxygen mass region, however, it turned out that for heavier nuclei adjustments of the interaction to specific mass regions are needed even when advanced many-body techniques are employed, so that the approaches lose their "ab initio" character. Such adjustments, in general, mean that the parameters of the interaction absorb implicitly some many-body dynamics, so that it becomes impossible to disentangle the effects related to the underlying "bare" interaction and those caused by the fermionic correlations. This, in turn, makes it difficult to compare theoretical models, in which these two ingredients are entangled in different ways. This problem can be viewed as the problem of variant expansions of the hypothetically exact Hamiltonian, or Lagrangian, based on different assumptions or on introducing different order parameters. 

In this work we discuss an approach which allows one formally to start with the bare nucleon-nucleon interaction and to take into account in-medium many-fermion correlations in a consistent and rather accurate way. As in the present formulation we consider only Hamiltonians with two-body interactions, the major quantity of interest to describe nuclear spectra is the in-medium two-fermion propagator in various channels. Respectively, the technique of choice is the Green function formalism and the equation of motion method which was extensively discussed in the past \cite{ST.16,DRS.98,AS.89}.  As follows from its formulation, the approach can be, if needed, straightforwardly generalized to three-body forces. 

It has been shown, for instance, in \cite{ST.16} that the usual hierarchy of equations for the many-body Green functions can be truncated at the two-fermion level and still contain many important long-range fermionic correlations. Such a truncation allows for a closed system of equations for two-fermion propagators, or response functions, which contains, in principle, all nuclear structure information related to one-body external operators. Recently we have established a connection of this approach with the class of approaches to the nuclear response derived by the time blocking method \cite{Tselyaev1989} which also takes two-fermionic correlations into account in a non-perturbative way.
The latter method, in its self-consistent implementation, is based on the relativistic meson-exchange nuclear QHD Lagrangian \cite{Ring1996} and extends the covariant response theory \cite{VretenarAfanasjevLalazissisEtAl2005,LiangVanGiaiMengEtAl2008} by effects of retardation induced by a strongly-correlated medium. The resulting approach restores approximately the time dependence neglected in the (relativistic) quasiparticle random phase approximation (RQRPA) \cite{PaarRingNiksicEtAl2003,NiksicMarketinVretenarEtAl2005} taking into account the most important  (resonant) effects of temporal non-localities, essential at the relevant excitation energies ($\sim$ 0-50 MeV).  In the original version of the relativistic time blocking approximation (RTBA) \cite{LitvinovaRingTselyaev2007} they are modeled by coupling of particle-hole pairs to collective vibrations within the "particle-hole plus phonon" (ph$\otimes$phonon) coupling scheme, also called particle-vibration coupling (PVC). The extended versions include like-particle superfluid pairing \cite{LitvinovaRingTselyaev2008} and quasiparticle-vibration coupling (QVC), two-phonon coupling \cite{LitvinovaRingTselyaev2010,LitvinovaRingTselyaev2013} and 
"two-quasiparticles plus N phonons" (2q$\otimes$Nphonon) configurations \cite{Litvinova2015}. The method has been applied successfully to excitation spectra of various closed and open-shell medium-mass spherical nuclei \cite{LitvinovaRingTselyaev2008,LRV.07,LitvinovaRingTselyaevEtAl2009,TPN.11,MassarczykSchwengnerDoenauEtAl2012,OEL.14,PFK.14,EgorovaLitvinova2016}. It has described well both high and low-energy parts of the spectra including the giant multipole resonances, the isospin splitting of the pygmy dipole mode \cite{EndresLitvinovaSavranEtAl2010,EndresSavranButlerEtAl2012,LanzaVitturiLitvinovaEtAl2014,NWL.16}, isoscalar dipole modes \cite{PellegriBraccoCrespiEtAl2014,KrzysiekKmiecikMajEtAl2016}, and stellar reaction rates of the r-process nucleosynthesis \cite{LitvinovaLoensLangankeEtAl2009}. The improvements in the description of excited states is related to the respective advancements in the single-particle sector \cite{LitvinovaRing2006,LitvinovaAfanasjev2011,AfanasjevLitvinova2015,Litvinova2016,KLRL.2017}.
Future developments of the relativistic nuclear response theory will be devoted to the generalized  relativistic quasiparticle time blocking approximation (RQTBA) \cite{Litvinova2015} which is shown to be more complete in the context of the equation of motion method in the sector describing the time evolution of two-body propagators \cite{LitvinovaSchuck2018}. In this contribution we discuss  the charge-exchange, or proton-neutron, version of the RQTBA (pn-RQTBA) and its applications to exotic nuclear systems.

%
\section{Equation of motion for the two-fermion propagator}
\label{EOM2}
\begin{figure}[h]
\centering
\sidecaption
\includegraphics[width=7cm,clip]{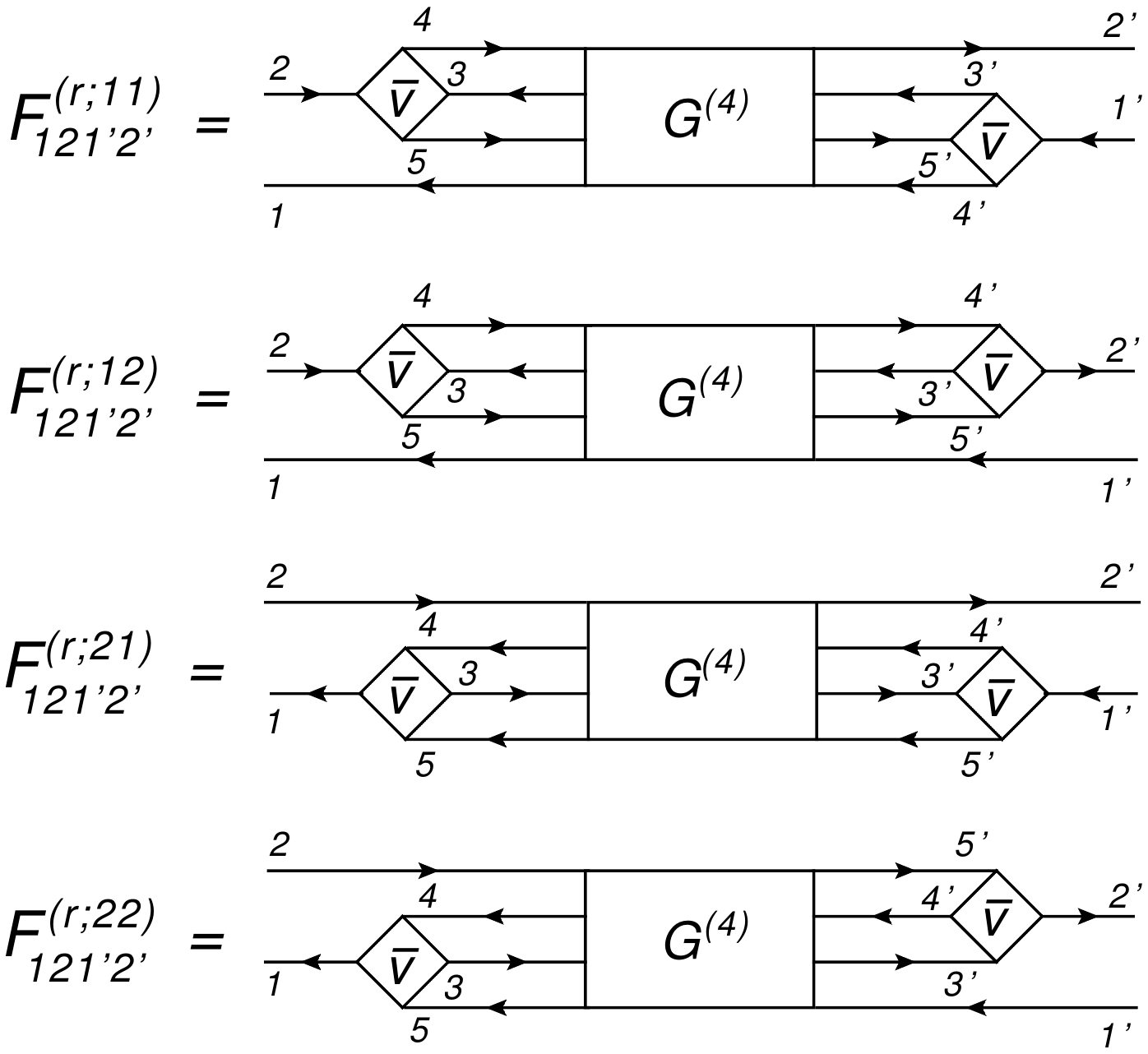}
\caption{Diagrammatic representation of the dynamical kernel $F^{(r)}_{12,1'2'}(t-t')$. Straight solid lines denote fermionic propagators, square block corresponds to the antisymmetrized nucleon-nucleon interaction $\bar v$, and the rectangular block $G^{(4)}$ is the two-particle-two-hole Green function (\ref{G4}).}
\label{fig-1}       
\end{figure}

Let us consider a many-fermion system without superfluidity. For such a system, 
the two-times two-fermion particle-hole propagator (response function) reads: 
\be
R(12,1'2') \equiv R_{12,1'2'}(t-t') = -i\langle T(\psi^{\dagger}_1\psi_2)(t)(\psi^{\dagger}_{2'}\psi_{1'})(t')\rangle = -i\langle T\psi^{\dagger}(1)\psi(2)\psi^{\dagger}(2')\psi(1')\rangle,
\label{phgf}
\ee
assuming $t_1 = t_2 = t, t_{1'} = t_{2'} = t'$. 
$T$ is the chronological ordering operator, $\psi(1),{\psi^{\dagger}}(1)$ are one-fermion (for instance, one-nucleon) fields:
\be
\psi(1) = e^{iHt_1}\psi_1e^{-iHt_1}, \ \ \ \ \ \ {\psi^{\dagger}}(1) = e^{iHt_1}{\psi^{\dagger}}_1e^{-iHt_1},
\ee
and the subscript '1' denotes the full set of the one-fermion quantum numbers in an arbitrary representation. 
The averaging in Eq. (\ref{phgf}) $\langle ... \rangle$ is performed over the ground state of the many-body system described by the Hamiltonian $H$:
\be
H = \sum_{12}{t}_{12}{\psi^{\dagger}}_1\psi_2 + \frac{1}{4}\sum\limits_{1234}{\bar v}_{1234}{\psi^{\dagger}}_1{\psi^{\dagger}}_2\psi_4\psi_3 = T + U
\ee
with the matrix elements of kinetic energy $t_{12} =  \delta_{12}\varepsilon_1$ and antisymmetrized interaction ${\bar v}_{1234} = v_{1234} - v_{1243}$ in the basis which diagonalizes $t_{12}$.
The equation of motion for the response function can be generated by taking the time derivative with respect to $t$ and a subsequent Fourier transformation of Eq. (\ref{phgf}). These operations lead to the following equation of the Dyson's type (in the operator form):
\be
R(\omega) = R^{(0)}(\omega) + R^{(0)}(\omega)W(\omega)R(\omega),
\label {Dyson2}
\ee
where the uncorrelated particle-hole propagator $R^{(0)}(\omega)$ is defined as:
\be
R^{(0)}_{12,1'2'}(\omega) = \delta_{11'}\delta_{22'}\frac{n_1 - n_2}{\omega - \varepsilon_{2} + \varepsilon_{1}}
\ee
with $n_1 = \langle{\psi^{\dagger}}_1\psi_1\rangle$ being the occupation number of the fermionic state $1$.
The integral kernel of Eq. (\ref{Dyson2}) is the Fourier transform of the following expression:
\be
F_{12,1'2'}(t-t') = \delta(t-t')
\langle [[V,{\psi^{\dagger}}_1\psi_2],{\psi^{\dagger}}_{2'}\psi_{1'}]\rangle + i\langle T[V,{\psi^{\dagger}}_1\psi_2](t)[V,{\psi^{\dagger}}_{2'}\psi_{1'}](t')\rangle, 
\label{F1}
\ee
more precisely, its irreducible part with respect to $R^{(0)}(\omega)$, so that $W(\omega) = F^{irr}(\omega)$. 
%
From Eq. (\ref{F1}) it is clear that the reducible kernel $F_{12,1'2'}(t-t')$ is separated into the instantaneous $F^{(0)}$ and the time-dependent $F^{(r)}$ (retarded or advanced) part:
\be
F_{12,1'2'}(t-t') = F^{(0)}_{12,1'2'}\delta(t-t') + F^{(r)}_{12,1'2'} (t-t'),
\label{F2}
\ee
and, obviously, the kernel $W(\omega)$ can be also decomposed so.
The instantaneous term represents the self-consistent phonon-mean field which is, thereby, derived from the bare interaction \cite{ST.16}.
The time-dependent part $F^{(r)}$ of the kernel $F$ is given by the mean value of a product of the two commutators and, after evaluating them, can be represented in the diagrammatic form given in Fig. \ref{fig-1},  where we have omitted the factors $\pm i/4$ in front of each diagram and
where a two-times two-particle-two-hole (4-fermion) Green function $G^{(4)}(543'1',5'4'31)$ is introduced according to:
\be
G^{(4)}(543'1',5'4'31) = \langle T({\psi^{\dagger}}_1{\psi^{\dagger}}_3\psi_5\psi_4)(t)
({\psi^{\dagger}}_{4'}{\psi^{\dagger}}_{5'}\psi_{3'}\psi_{1'})(t')\rangle. 
\label{G4}
\ee
Thus, one can see that the dynamical kernel of Eq. (\ref{Dyson2}) requires the knowledge about the two-times two-particle-two-hole (four-body) propagator $G^{(4)}$ of Eq. (\ref{G4}). The EOM for this propagator can be generated analogously, however, it includes higher-rank propagators, resulting in an infinite chain of EOM's. A very reasonable and efficient truncation scheme was proposed in Ref. \cite{ST.16}, where $G^{(4)}$ is decomposed in its fully uncorrelated part, and a correlated part containing terms with products of non-correlated and correlated two-fermion propagators (response functions) of the type (\ref{phgf}). In this way, the problem of computing the two-fermion propagator reduces to the closed system of equations:
\be
{\hat R}(\omega) = {\hat R}^{(0)}(\omega) + {\hat R}^{(0)}(\omega)W[{\hat R}(\omega)]{\hat R}(\omega),
\label{Dyson3}
\ee
where we complement the particle-hole propagator considered above by the particle-particle, hole-hole and, formally, hole-particle ones defined analogously: \be
{\hat R} = \Bigl\{ R^{(ph)}, R^{(hp)}, R^{(pp)}, R^{(hh)} \Bigr\}, \ \ \ \ \ \ \ {\hat R}^{(0)} = 
\Bigl\{ R^{(0; ph)}, R^{(0; hp)}, R^{(0; pp)}, R^{(0; hh)} \Bigr\}.
\label{scresponse}
\ee
Remarkably enough, in this approximation all frequency (time) dependence of the kernel originates from the internal two-fermion propagators, or response functions, which are themselves the main variables. Notice here that ${\hat R}$ is, in principle, fully defined by the bare nucleon-nucleon interaction. In the practical calculations, however, one needs a reasonable initial approximation for ${\hat R}$ to start the iterative procedure of solving Eq. (\ref{Dyson3}). %
\begin{figure}[h]
\centering
\sidecaption
\includegraphics[width=7cm,clip]{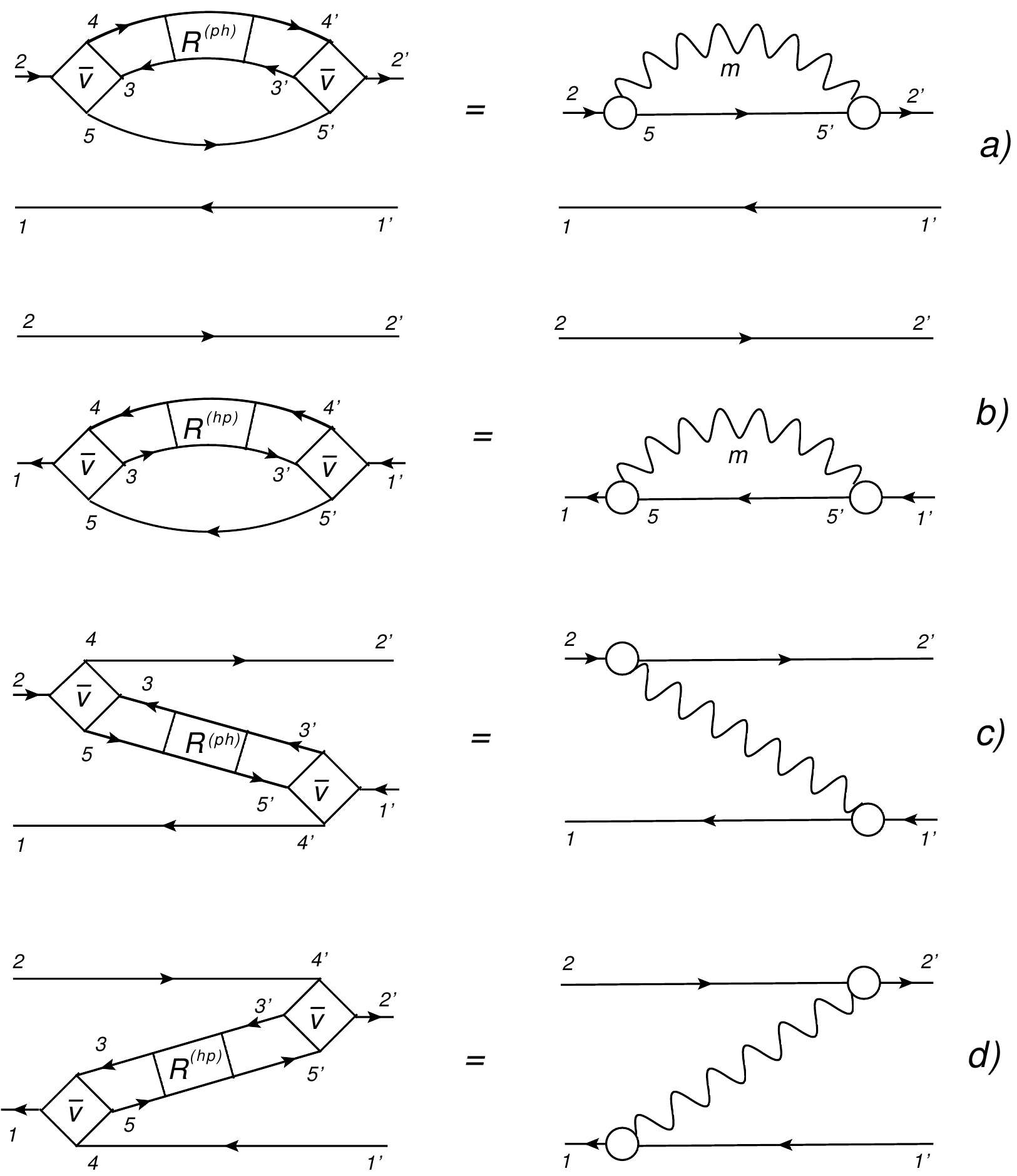}
\caption{Diagrammatic mapping of the single-correlated terms of $F^{(r)}$ containing $R^{(ph)}$ to the PVC kernel. Circles stand for the QVC vertices, wavy lines denote the phonon propagators, and the rectangular blocks represent the particle-hole and the hole-particle response functions.}
\label{fig-1a}       
\end{figure}
One of the possibilities is to calculate the initial ${\hat R}$ being confined by only the static part of the integral kernel of Eq. (\ref{F2}). This idea was realized in the RQTBA approach where this static part is approximated by the exchange of effective mesons whose coupling constants and masses are adjusted to the nuclear ground state properties on the Hartree level \cite{VretenarAfanasjevLalazissisEtAl2005}. The dynamical part of the kernel was modeled by the (quasi)particle vibration coupling in the lowest order with respect to the phonon coupling vertex, following the non-relativistic (quasiparticle) time blocking approximation \cite{Tselyaev1989,LitvinovaTselyaev2007,Tselyaev2007}. In the relativistic framework this approach has, thereby, become fully self-consistent and parameter-free using only the parameters of the covariant density functional theory \cite{LitvinovaRingTselyaev2007,LitvinovaRingTselyaev2008}. Although up until now in the practical implementations RQTBA starts from the relativistic RQRPA for the ${\hat R}(\omega)$ and performs only the first iteration of Eq. (\ref{Dyson3}), it gives a tremendous improvement compared to RQRPA and a very good description of experimental data, which indicates fast convergence of the iterations of Eq. (\ref{Dyson3}) in this scheme.

Fig. \ref{fig-1a} reveals to which extent the dynamical kernel of RQTBA includes the long-range correlations contained in $G^{(4)}$. The leading terms with the minimal (second-order) PVC are plotted on the right hand side and their exact mapping to the terms of the EOM kernel is shown. One can see that the original PVC-RQTBA includes only terms with a single particle-hole correlation function and neglect those with two correlated particle-hole (or particle-particle and hole-hole) lines. These contributions were included in the generalized RQTBA \cite{Litvinova2015}. Its numerical implementation, however, is not completed yet although the first step in this direction has been done in \cite{Tselyaev2007,LitvinovaRingTselyaev2010,LitvinovaRingTselyaev2013}. More details about the content of PVC-RQTBA in the framework of the EOM method can be found in Ref. \cite{LitvinovaSchuck2018}.



\section{Applications to isospin-flip excitations} 
\label{applications}
\begin{figure}[h]
\centering
\includegraphics[width=14.5cm,clip]{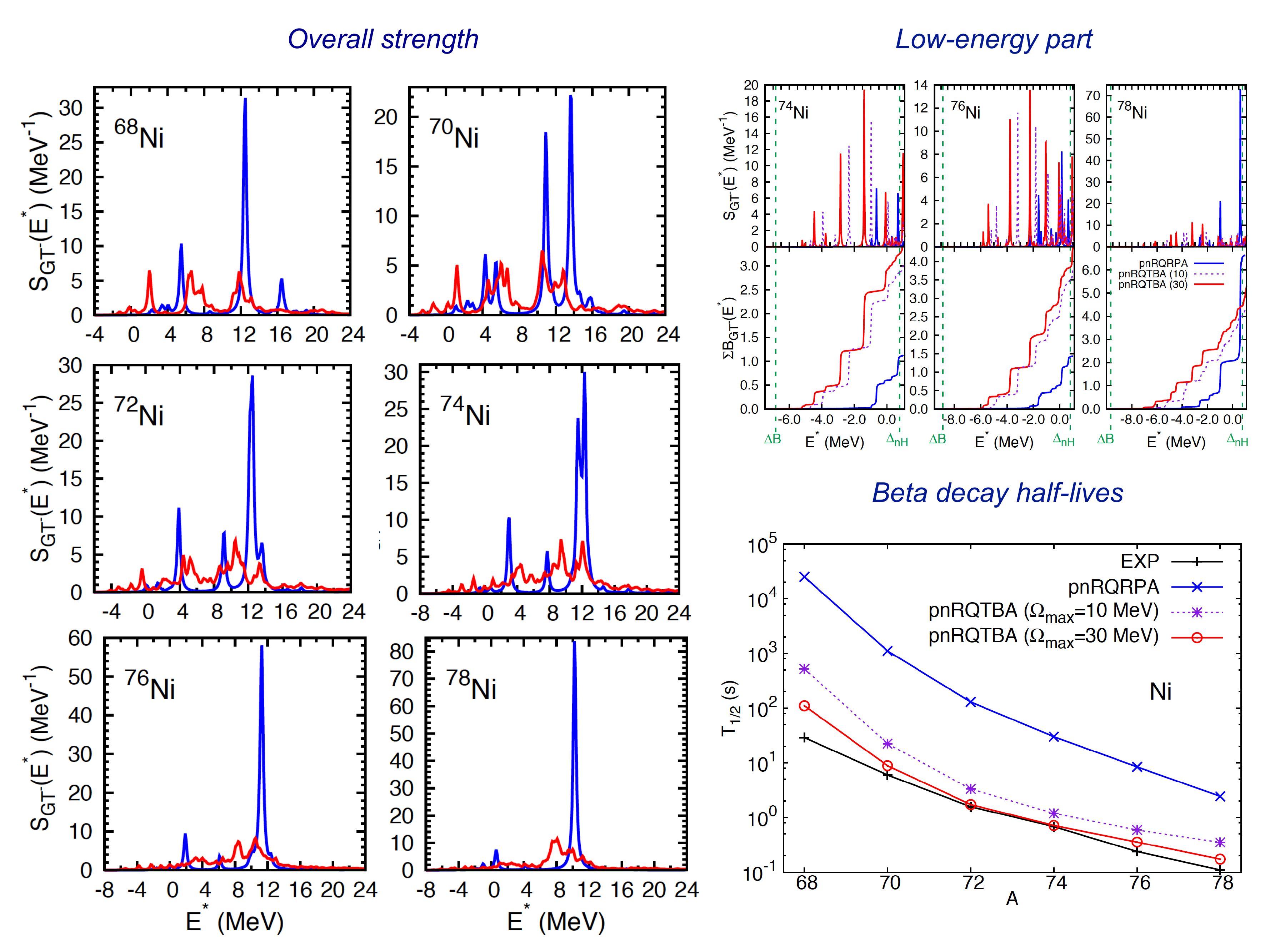}
\caption{(Adopted from Ref. \cite{RobinLitvinova2016}) Gamow-Teller strength distributions and beta decay half-lives in $^{68-78}$Ni. Left: the overall GTR strength distribution $^{68-78}$Ni; top right: its low-energy fraction in $^{74-78}$Ni in the Q$_{\beta}$ window; bottom right: the beta decay half-lives.  Blue curves and symbols show the results obtained in pn-RQRPA, the red color is used to present the pn-RQTBA calculations, and the purple dashed curves and symbols represent the pn-RQTBA calculations in a restricted phonon space.
See Ref. \cite{RobinLitvinova2016} for details.}
\label{fig-2}       
\end{figure}

The RQTBA turned out to be rather successful in describing the nuclear excitations of the neutral type in both high and low energy regions \cite{LitvinovaRingTselyaev2008,LitvinovaRingTselyaevEtAl2009,MassarczykSchwengnerDoenauEtAl2012,EndresLitvinovaSavranEtAl2010,EndresSavranButlerEtAl2012,LanzaVitturiLitvinovaEtAl2014,EgorovaLitvinova2016,PellegriBraccoCrespiEtAl2014,KrzysiekKmiecikMajEtAl2016}. Therefore, it was interesting to adopt the method to the charge-exchange channels. This has been done first for closed-shell nuclei \cite{MLVR.12,LBFMZ.14} and recently generalized for superfluid systems in Ref. \cite{RobinLitvinova2016}. 
Fig. \ref{fig-2} illustrates the performance of the latter approach, which is named the proton-neutron RQTBA (pn-RQTBA), for the Gamow-Teller strength distribution in the chain of neutron-rich nickel isotopes $^{68-78}$Ni: the overall $\beta^-$ GTR strength distribution in $^{68-78}$Ni, its low-energy fraction in $^{74-78}$Ni and the beta decay half-lives of $^{68-78}$Ni.
The overall GTR strength distributions (left) show the effect of the quasiparticle vibration coupling which is included  consistently in pn-RQTBA and neglected in pn-RQRPA. A strong fragmentation of the strength and its spread to a wider energy region is the major effect which occurs also in neutral channels. This calculation was performed with 200 keV value for the smearing parameter (which is within the typical range of the continuum width for medium-mass nuclei). A more detailed illustration of this spread to the low-energy region is given in the right top panels, where the strength was calculated with 20 keV smearing in order to resolve the individual states. The $Q_{\beta}$ energy window is outlined by the green dashed lines. Below (right bottom panel) the corresponding beta decay half-lives are displayed. The more strength is found in the $Q_{\beta}$ window, the more probable is the beta decay and, respectively, the shorter is the lifetime of the nucleus. This correlation is not direct because of the lepton phase space effects, however, one observes a systematic decrease of the half-lives  by one or two orders of magnitude when the QVC effects are included. The red curves and symbols correspond to pn-RQTBA calculation with the larger phonon space (phonons with up to 30 MeV excitation energies and up to J=6 multipolarities) while the purple ones show the results obtained with 10 MeV energy cutoff for the phonon modes. In this way, the importance of the completeness of the phonon space is illustrated. Notice here that no static proton-neutron pairing with free  parameters was employed in these calculations.


In Ref. \cite{28si} the ($^{10}$Be,$^{10}$B[1.74 MeV]) charge-exchange reaction at 100 AMeV was presented as a new probe,
which is capable of isolating the isovector ($\Delta$T = 1) non-spin-transfer ($\Delta$S = 0) nuclear response. 
The N=Z neutron deficient nucleus $^{28}$Si was chosen for this study at the National Superconducting Cyclotron Laboratory (NSCL).  
A secondary $^{10}$Be beam produced by fast fragmentation of $^{18}$O nuclei at the NSCL Coupled Cyclotron Facility, the dispersion-matching
technique with the S800 magnetic spectrometer, and the high-precision $\gamma$-ray tracking with the Gamma Ray Energy Tracking Array (GRETINA)
were used to obtain a clean $\Delta$S = 0 excitation-energy spectrum in $^{28}$Al. Monopole and dipole contributions were extracted through a multipole decomposition analysis, and, thereby, the isovector giant dipole (IVGDR) and the isovector giant monopole (IVGMR) resonances were identified. The results show that this probe is a powerful tool for studying the elusive IVGMR, which is of interest for performing a stringent test of theoretical approaches at high excitation energies and for constraining the bulk properties of nuclei and nuclear matter. 
\begin{figure}[h]
\centering
\sidecaption
\includegraphics[width=7cm,clip]{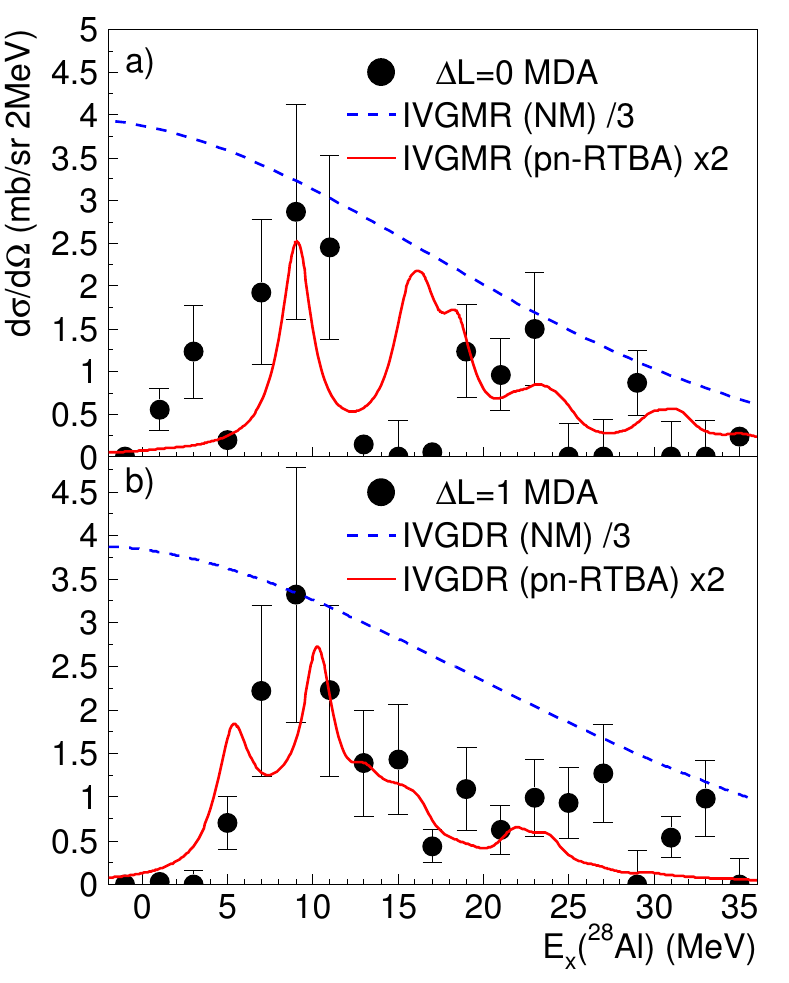}
\caption{(Reprinted from Ref. \cite{28si}) The experimental isovector giant monopole (top) and isovector giant dipole (bottom) resonances in $^{28}$Si extracted from $^{28}$Si($^{10}$Be,$^{10}$B)$^{28}$Al charge-exchange reaction (black circles with error bars) compared to the normal-modes (NM) calculation (blue dashed curves) and to the cross sections computed with the pn-R(Q)TBA strength functions (red solid curves). The details are given in Ref. \cite{28si}.}
\label{fig-3}       
\end{figure}
Fig. \ref{fig-3} shows the extracted distributions compared with
theoretical calculations based on the normal-modes (NM) formalism and on the pn-R(Q)TBA strength functions. One can see that pn-R(Q)TBA describes very reasonably the shapes of the experimental cross sections, which can not be achieved within pn-R(Q)RPA. The latter emphasizes the importance of long-range correlations for these isospin-flip resonances. However, even the pn-R(Q)TBA
gives overall smaller values, compared to the measured cross sections. The latter is most likely caused by the deficiencies in the reaction calculations based on the distorted-wave Born approximation. 

Another recent application of pn-RQTBA is the spin-isospin excitations in $^{100}$Nb, which were studied by the charge-exchange $^{100}$Mo(t, $^3$He) reaction at NSCL \cite{100mo}.  The neutron decays from the excited $^{100}$Nb were also observed. The statistical and direct decay branches were both identified in the spectra. The upper limit for the direct-decay branching ratio was determined to be 20$\pm$6\%, which revealed the decay predominantly happened via the statistical process.
\begin{figure}[h]
\centering
\sidecaption
\includegraphics[width=7cm,clip]{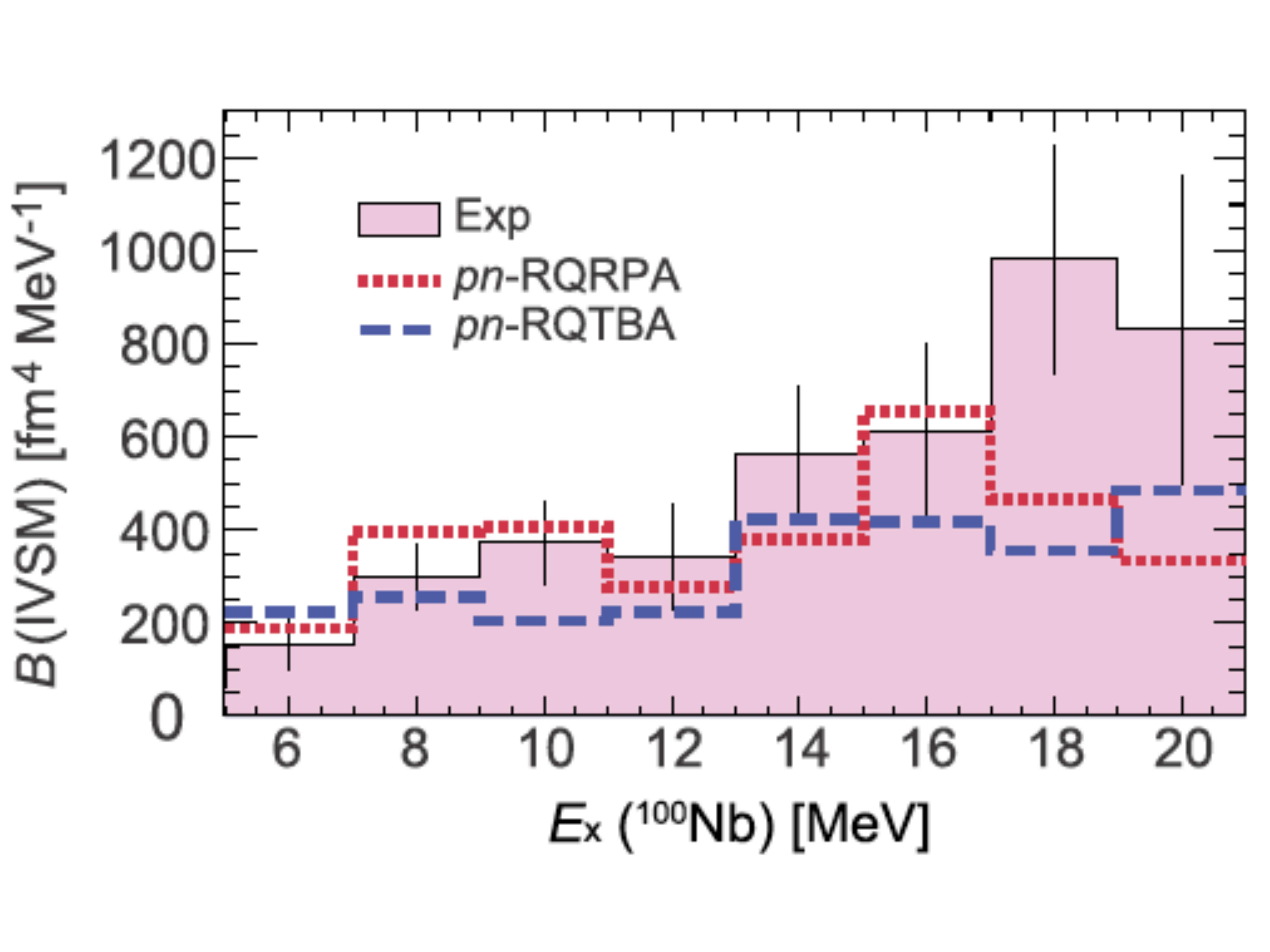}
\caption{(Reprinted from Ref. \cite{100mo}) The isovector spin monopole strength distribution. The black
histograms are the experimental results and  the error bars contain only the statistical contributions. The pn-RQRPA and pn-RQTBA calculations are shown by dashed red and the dashed blue histograms, respectively. See Ref. \cite{100mo} for details.}
\label{fig-4}       
\end{figure}
%
%
%
%
Fig. \ref{fig-4} presents the isovector spin monopole (IVSM) resonance  extracted from this measurement, together with the theoretical pn-RQRPA and pn-RQTBA calculations in the form of histograms consistent with the experimental energy resolution. 
The IVSM resonance is the overtone of the Gamow-Teller resonance: their operators differ by the radial formfactors only. In practice, the IVSM resonance reflects the leading-order effect of the momentum transfer dependence of the spin-isospin response dominated by the GTR.This effect was studied in detail, for instance, in Ref. \cite{MMPD.12}. It has been shown, in particular, that the IVSM resonance can absorb a few percent of the total Ikeda sum rule and, thereby, contribute to the quenching of the GTR. Thus, understanding the formation of the IVSM resonance is of a great importance because the identification of microscopic mechanisms of the GTR's  quenching remains one of the most difficult unsolved problems in the nuclear structure physics. One can see from the Fig. \ref{fig-4} that both pn-RQRPA and pn-RQTBA give reasonable strength distributions, which means that for the IVSM excitations the long-range correlations are most likely less important than for the non-spin-flip isovector monopole resonance discussed above. The discrepancies between the data and calculations can be, therefore, attributed to possible deformation effects or to deficiencies of the static interaction, such as the overall simplicity of its one-boson-exchange character and the absence of the delta meson. This points to further work in those directions. 
\section{Summary}
We discussed recent developments of relativistic nuclear field theory which has recently been put in the context of the general equation of motion method. This has allowed for linking the methods developed previously for the nuclear response function, which are based on the effective interaction, with the underlying  bare nucleon-nucleon forces. At the same time, the approach to the nuclear response function developed in the framework of the time blocking method has been related to the systematic EOM treatment, which allows one to evaluate explicitly the contributions neglected in the RQTBA and to improve it considerably. 

Recent developments and applications of the charge-exchange RQTBA have been discussed. First of all, pairing correlations have been taken into account self-consistently in this channel, so that systematic investigations of spin-isospin excitations and beta decay in open-shell nuclei are now possible. The advantage of the nuclear response theory in the charge-exchange channel is the possibility to separate (approximately, but rather accurately) the contributions from the pion exchange and the $\rho$-meson exchange and, thereby, get cleaner constraints on their coupling strengths. In particular, the unnatural-parity probes, such as the GTR, give the constraints on the pion exchange, and the excitations with natural parity on the $\rho$-meson exchange interaction. The shapes of the spectra come to a better agreement with data when dynamical, or long-range, correlations are taken into account, however, in some cases the comparison with data points to the necessity of further improvements in this sector.

\section*{Acknowledgements}
The authors greatly appreciate discussions with P. Ring and R.G.T. Zegers. Special thanks to T. Marketin for providing a part of the code for pn-RRPA matrix elements. This work is partly supported by US-NSF grant PHY-1404343, NSF Career grant PHY-1654379, by the Institute for Nuclear Theory under US-DOE Grant DE-FG02-00ER41132 and by JINA-CEE under US-NSF Grant PHY-1430152.

\end{document}